# Symmetry of two-terminal, non-linear electric conduction

A. Löfgren[a], C.A. Marlow[b], I. Shorubalko[a], R.P. Taylor[b], P. Omling[a], L. Samuelson[a], and H. Linke[b,*]

[a] *Solid State Physics, Lund University, Box 118, S – 22100 Lund, Sweden*
[b] *Physics Department, University of Oregon, Eugene OR 97403-1274, USA*

The well-established symmetry relations for linear transport phenomena can not, in general, be applied in the non-linear regime. Here we propose a set of symmetry relations with respect to bias voltage and magnetic field for the non-linear conductance of two-terminal electric conductors. We experimentally confirm these relations using phase-coherent, semiconductor quantum dots.

Symmetries with respect to the sign of a bias voltage and the direction of an applied magnetic field, $B$, are central to our understanding of electron transport phenomena. In the linear response regime, the Onsager-Casimir relations, $\sigma_{\alpha\beta}(B) = \sigma_{\beta\alpha}(-B)$, describe these symmetries in terms of the local conductivity tensor [1]. These relations were derived for macroscopic, disordered solid-state conductors where the conductor boundaries are unimportant. In mesoscopic samples, the characteristic length scales for elastic and inelastic (phase-breaking) scattering can exceed the dimensions of the device. In this limit a local description of transport is not possible, and the reciprocity theorem, $R_{12,34}(B) = R_{34,12}(-B)$, must be used [2, 3]. For two-terminal conductors, the reciprocity theorem reduces to $G_{12}(B) = G_{12}(-B)$, where $G_{12}$ is the conductance with the current flowing from contact 1 to 2. The sign of the source-drain bias voltage and the orientation of the measurement leads are of no consequence in the linear response regime, such that $G_{12}(B) = G_{21}(B)$.

The reciprocity theorem breaks down in the non-linear response regime [4-7]. In the general case, where the conductor has no symmetry (e.g. due to disorder), $G_{12}(V) \neq G_{12}(-V)$. This is because, if an applied voltage modifies the asymmetric device potential, the resulting device potential depends on the voltage sign [6-8]. Similarly, no symmetries with respect to magnetic field are expected for an asymmetric device, that is, $G_{12}(V, B) \neq G_{12}(V, -B)$ (for an illustration, see Fig. 1).

While the general breakdown of the reciprocity theorem in the non-linear transport regime is well known [4-7], a systematic evaluation of surviving symmetries in this regime has not previously been attempted. It is the point of the present paper to establish a complete set of symmetry relations for the non-linear conductance of two-terminal conductors. One important motivation is that the non-linear regime is fundamental to applications of sub-micron electronic devices, for which linear response is limited to very small voltages [4, 5].

In the following we consider conductors without significant disorder. We first propose a set of general symmetry relations with respect to bias voltage and magnetic field for the non-linear electric conductance. We will show that symmetry of non-linear transport requires geometrical symmetry of the conductor − a substantial experimental challenge in terms of fabrication and material quality. Using purposely-designed semiconductor quantum dots, we then demonstrate that the symmetry relations are experimentally observed, and that deviations from perfect geometrical symmetry can be measured.

Without loss of generality we consider triangular conductors because of their simple geometrical shape. We refer to a device as left-right (LR) symmetric when it possesses a symmetry axis perpendicular to the current direction, and up-down (UD) symmetric when it possesses a symmetry axis parallel to the current direction. We consider the symmetry of the non-linear electrical conductance under reversal of voltage, magnetic field, and lead orientation. In this context it is important to note that in a real experimental set-up the reversal of voltage ($V \to -V$) is not generally equivalent to physically interchanging the leads attached to the probes ($G_{12} \to G_{21}$), because the circuit used to measure the conductance may itself be asymmetric. For instance,

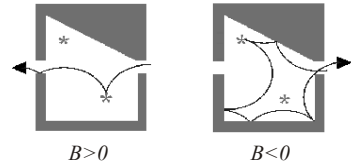

$B>0$        $B<0$

FIG. 1. Schematic electron trajectories for positive and negative magnetic fields. In the absence of symmetry in a mesoscopic device, the conductance is not expected to be symmetric with respect to the direction of a magnetic field, when a bias voltage defines a source and a drain contact.

the gate voltage $V_g$ used to electrostatically define the conductor's shape is usually set with respect to the drain contact on one side of the device, breaking the circuit symmetry. When appreciable source-drain voltages are used, the resulting gradient in the local electro-chemical potential along the conductor deforms the device potential defined by the gate in a way that depends on the voltage sign. This can lead to circuit-induced asymmetry (CIA) of the conductance even when the device itself is LR-symmetric [9-11]. In order to avoid CIA, special care must be taken in the device design [12]. Here we focus on so-called "rigid" devices in which CIA is not significant, and refer the reader to Ref. [11] for a discussion of devices that are not rigid.

For rigid devices, regardless of their symmetry, a voltage reversal is equivalent to swapping source and drain leads, such that

$$G_{12}(V, B) = G_{21}(-V, B) \quad \text{(rigid)} \quad (1)$$

This relation is illustrated in Fig. 2 (compare, for instance, configurations A and G or D and F).

For the special case of rigid devices that are LR-symmetric we expect

$$G_{12}(V, B) = G_{12}(-V, -B) \quad \text{(LR, rigid)} \quad (2)$$

Eq. (2) holds independent of whether or not the device is UD-symmetric, but is not expected if LR-symmetry is absent. This can be seen by comparing, for instance, A and D or B and C in Fig. 2.

UD-symmetry implies that, for a given voltage, reversal of a magnetic field perpendicular to the device plane should be of no consequence for electron transport [8]:

$$G_{12}(V, B) = G_{12}(V, -B) \quad \text{(UD)} \quad (3)$$

This can be seen by comparing, for instance, A and B in Fig. 2. Note, however, that the absence of LR-symmetry implies that, in the non-linear regime $G_{12}(V) \neq G_{12}(-V)$, regardless of the magnetic field sign [5, 7, 8]. Eq. (3) does not involve a reversal of lead orientation or voltage sign and is therefore valid for both rigid and non-rigid devices.

Finally, we note that the conductance of a LR-symmetric device (regardless of rigidity and UD-symmetry) is expected to be invariant upon reversal of lead orientation and of the external magnetic field:

$$G_{12}(V, B) = G_{21}(V, -B) \quad \text{(LR)} \quad (4)$$

Relationships (1) – (4), the first main result of our paper, are based on fundamental symmetry arguments and are therefore expected to hold in both the classical and quantum regimes of transport.
In order to test these relationships, we used ballistic semiconductor devices defined by deep wet etching in modulation-doped, 9 nm thick InP/GaInAs. The devices were of equilateral-triangular shape with a side length of 1 μm, smaller than the electrons' elastic mean free path of 6.1 μm and smaller than the phase-coherence length $l_\phi$ = 3.5 μm at $T$ = 230 mK and $V$ = 0 ($l_\phi$ = 1.7 μm at $T$ = 230 mK and $V$ = 3 mV). In this phase-coherent regime of electron transport, the wave-like nature of the carriers leads to conductance fluctuations (CF) as a function of an applied magnetic field. Because of their origin in wave-interference, and because of the short Fermi wave-length (30 nm), details of the CF are known to be sensitive to the exact shape of the potential forming the device and to defects or impurities [13]. Phase-coherent measurements of CF are therefore particularly well suited to test the influence of device geometry and of disorder on the conductance symmetry. Contact openings used to measure the conductance were positioned such that either UD-symmetric (Fig. 3) or LR-symmetric (Fig. 4) quantum dots were formed. Two-terminal magneto-conductance measurements were carried out in four-point geometry. A small ac signal (rms amplitude 20 μV, comparable to $kT \approx 20$ μeV) was added to a tunable dc bias voltage $V$. The differential conductance $g_{ij} = dI_{ij}/dV_{ij}$ was measured using lock-in techniques in order to reduce measurement noise. We checked that there was no significant non-ohmic behavior in the circuit.

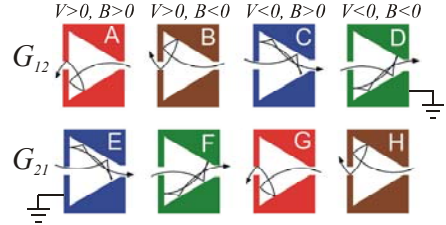

FIG. 2. Illustration of the symmetry relations expected for a rigid device in the non-linear regime and at finite magnetic field. The upper and lower rows show the two possible lead configurations $G_{12}$ and $G_{21}$, distinguished by the position of the grounding point relative to the device. Different classical electron trajectories illustrate the difference in transmission probability that results when the potential depends on the sign of the voltage applied to the source contact. Positive magnetic field is taken to be into the page.

Fig. 3 shows $g_{ij}$ for a bias voltage $|V|$ = 1 mV $\approx 50\ kT/e$ as a function of a perpendicular magnetic field for an UD-symmetric, triangular quantum dot [14]. The eight traces shown are individual measurements taken over the course of two days in the eight possible configurations of sign of the bias voltage, direction of the magnetic field, and lead orientation (see Fig. 2). As expected for a device lacking LR symmetry, the CF are not symmetric in $V$, as we note from a comparison of Figs. 3(a) and 3(b), or Figs. 3(c) and 3(d) [15]. However, for a rigid device, Eq. (1) predicts that reversal of the leads and bias voltage should lead to identical CF, regardless of the device symmetry. The similarity of traces shown in the same color (for instance A and G, or D and F) qualitatively verifies Eq. (1), and shows that the device used in Fig. 3 can be regarded as rigid [12].

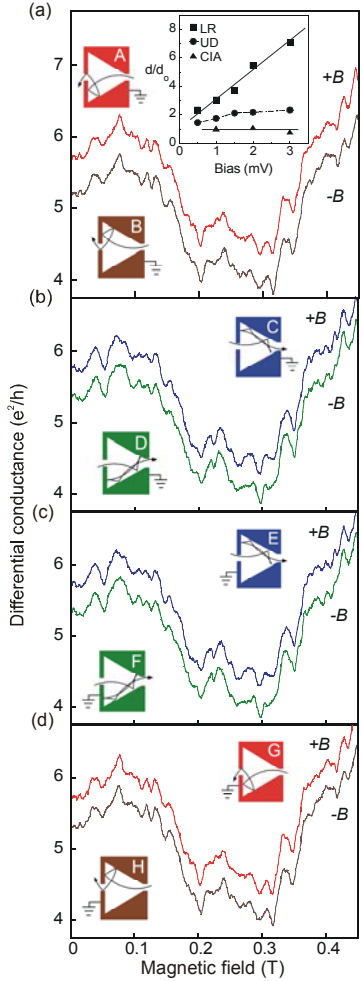

FIG.3. Magneto-conductance fluctuations for an UD-symmetric quantum dot measured in the eight possible different configurations of lead orientation, sign of bias voltage, and sign of magnetic field (capital letters refer to the panels in Fig. 2): (a) shows $g_{12}(V = +1$ mV, $\pm B)$, (b) shows $g_{12}(V = -1$ mV, $\pm B)$, (c) shows $g_{21}(V = +1$ mV, $\pm B)$, and (d) shows $g_{21}(V = -1$ mV, $\pm B)$. The lower trace in each panel has been offset by $-0.5$ $e^2/h$ for clarity. The inset to (a) shows $d_{LR}/d_0$, $d_{UD}/d_0$, and $d_{CIA}/d_0$ as a function of $V$ (lines are guides to the eye).

According to Eq. (3), in the presence of perfect UD symmetry conductance fluctuations should be unaltered when the direction of the magnetic field is reversed. This prediction can be tested by comparing the pairs of traces in the individual panels in Fig. 3 (e.g. A and B or C and D). Again, striking similarities are observed.

In order to quantify the difference between two magneto-conductance traces, say the difference $d_{AB}$ between traces $g_A(B)$ and $g_B(B)$ measured in configurations A and B, respectively, we determine the root mean square (rms) of their difference, using $10^3$ data points spaced by 0.5 mT between $B = 0$ and $B^{max} = \pm 0.5$ T:

$$d_{AB} = \sqrt{\frac{1}{B^{max}} \int_0^{B^{max}} [g_A(B) - g_B(B)]^2 dB} \qquad (5)$$

The value $d_{AB} = 0$ would correspond to identical traces. To calibrate the influence of experimental noise and setup instabilities on $d$ we use two CF traces recorded two days apart in nominally identical configurations ($V = 0$). Separately evaluating $d$ for the traces recorded for positive and negative magnetic field and then averaging the results, we find $d_0 = 2.99\times10^{-2}$ $e^2/h$, a value comparable to experimental noise ($\approx 0.5$ %) of the device conductance. In comparison, the four pairs of traces that should be identical if the device is rigid (A − G, B − H, C − E, D − F) yield an averaged value of $d_{CIA} = (d_{AG} + d_{BH} + d_{CE} + d_{DF})/4 = 3.10\times10^{-2}$ $e^2/h$ and $d_{CIA}/d_0 = 1.04$. A comparison of data sets that, according to Eq. 2, should be the same if the device was LR symmetric (A − D, B − C, E − H, F − G), yields $d_{LR}/d_0 = 3.05$. In comparison, a test for UD symmetry (A − B, C − D, E − F, G − H) yields the averaged value $d_{UD}/d_0 = 1.77$. In other words, the intentional absence of LR-symmetry in the device geometry causes the largest conductance asymmetry, while unintentional deviations from UD-symmetry, such as material and fabrication imperfections, have a significantly smaller, but measurable effect. The effect of CIA in our devices is not significant compared to experimental noise, confirming that the device is rigid. Note, however, that CIA can be substantial in other devices, for instance in some surface-gated devices [9, 10, 12].

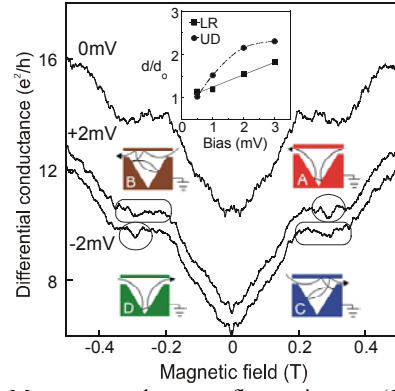

FIG. 4. Magneto-conductance fluctuations $g_{12}(V, B)$ of a LR-symmetric device for $V = 0$, $V = +2$ mV and $V = -2$ mV (capital letters in each measurement configuration refer to the corresponding panel in Fig. 2). Data are offset for clarity. Note that $g_{12}(V, B) \neq g_{12}(V, -B)$, while $g_{12}(V, B)$ and $g_{12}(-V, -B)$ show very similar features, as predicted by Eq. (2) (see, e.g. the marked features). Inset: $d_{LR}/d_0$ and $d_{UD}/d_0$ as a function of $V$ ($d_0 = 4.21\times10^{-2}$ $e^2/h$ for this device). Lines are guides to the eye.

The inset to Fig. 3(a) shows the quantified asymmetries (normalized to $d_0$) as a function of increasing bias voltage. Consistent with a first order non-linear effect, $d_{LR}$ increases approximately linearly with bias voltage. On the other hand, $d_{UD}$, which is attributed to imperfections in the UD-symmetry of the device, which are not expected to change with voltage, increases only weakly with $V$. At all voltages used, the influence of CIA remained insignificant compared to the noise level ($d_{CIA}/d_0 \approx 1$).

For comparison with the UD-symmetric device discussed so far, in Fig. 4 we show CF for the LR-symmetric device. Whereas at $V = 0$ (linear regime) the conductance is symmetric in $B$, at finite $V$ (non-linear regime) each of the two data traces taken is not symmetric in $B$, due to the absence of UD symmetry. However, one can see by comparing the marked conductance features that $g_{12}(V, B) \approx g_{12}(-V, -B)$. This observation confirms Eq. (2) and indicates that the device is rigid, consistent with our conclusion about the UD-symmetric device. We therefore expect that any $d_{LR}$ observed should be due to unintentional deviations from LR symmetry. Indeed, at all bias voltages $d_{LR}/d_0$ for this device (see inset to Fig. 4) is substantially smaller than for the LR-asymmetric device used in Fig. 3. As one would expect intuitively from the symmetry of the device, for small bias $d_{LR}$ is also smaller than $d_{UD}$ (inset to Fig. 4). Note, however, that the values found for $d_{UD}/d_0$ in the UD and the LR-symmetric devices are comparable, highlighting an interesting open question: At present, no theoretical prediction about the dependence of $d_{UD}$ on disorder, magnetic field, or bias voltage is available. Our data (see insets to Fig. 3(a) and Fig. 4) suggest a sub-linear increase of $d_{UD}$ with $V$, and little sensitivity to the amount of intentional asymmetry.

The symmetry relations demonstrated here were predicted based on symmetry arguments. A natural next step would be a rigorous theoretical study along the lines of Ref. [3], and applicable to the non-linear regime of transport.

*Acknowledgments*. P.E. Lindelof for useful discussions, I. Maximov for lithography and W. Seifert for crystal growth. Supported by an NSF IGERT (C.A.M.), an NSF CAREER award (H.L.), a Cottrell scholarship (R.P.T), the ONR, the Swedish Foundation for Strategic Research, and the Swedish Research Council.
*Corresponding author: linke@uoregon.edu